\documentclass[aps,prd,showkeys,nofootinbib,notitlepage,longbibliography,twocolumn, superscriptaddress]{revtex4-1}
\usepackage[utf8]{inputenc}
\usepackage{amsmath,amsthm,amsfonts,amssymb,amscd}
\usepackage{multirow,booktabs}
\usepackage[table]{xcolor}
\usepackage{fullpage}
\usepackage{siunitx}
\usepackage{lastpage}
\usepackage{enumitem}
\usepackage{fancyhdr}
\usepackage{mathrsfs}
\usepackage{wrapfig}
\usepackage{ulem}
\usepackage{setspace}
\usepackage{calc}
\usepackage{cancel}
\usepackage{graphicx,bm}
\usepackage{appendix}
\usepackage[colorlinks,linkcolor=blue,citecolor=blue]{hyperref}
\date{\today}
\newcommand{\dd}{\mathrm{d}}
\begin{document}
\title{Spacetime dynamics of chiral magnetic currents in a hot non-Abelian plasma}
\author{Sebastian Grieninger}
\email{sebastian.grieninger@stonybrook.edu}
\affiliation{Center for Nuclear Theory, Department of Physics and Astronomy,
Stony Brook University, Stony Brook, New York 11794–3800, USA}

\author{Dmitri E. Kharzeev}
\email{dmitri.kharzeev@stonybrook.edu}
\affiliation{Center for Nuclear Theory, Department of Physics and Astronomy,
Stony Brook University, Stony Brook, New York 11794–3800, USA}
\affiliation{Department of Physics, Brookhaven National Laboratory
Upton, New York 11973-5000, USA}
\begin{abstract}The correlations of electric currents in hot non-Abelian plasma are responsible for the experimental manifestations of the chiral magnetic effect (CME) in heavy-ion collisions. We evaluate these correlations using holography, and show that they are driven by large-scale topological fluctuations. In a non-Abelian plasma with chiral fermions, local axial charge can be generated either by topological fluctuations (creating domains with nonzero Chern-Simons number) or by thermal fluctuations. Within holography, we investigate the dynamical creation of the axial charge and isolate the imprint of the topological dynamics on the spatial correlations of electric current. In particular, we show that the spatial extent of the current correlation is quite large ($\sim 1\  {\rm fm}$) and grows with time, which is consistent with sphaleronlike dynamics. We provide numerical estimates for this spatial size that can be used as an input in phenomenological analyses.\end{abstract}
\maketitle

\section{Introduction}Non-Abelian gauge theories, including Quantum Chromodynamics (QCD), possess topologically nontrivial configurations of gauge fields \cite{Belavin:1975fg}. The topology of these gauge fields is characterized by Chern-Simons number \cite{Chern:1974ft} which differentiates between energetically degenerate but topologically distinct ground states. Tunneling transitions between such energy-degenerate vacuum sectors are mediated by instantons, classical Euclidean  solutions to the Yang-Mills field equations \cite{Belavin:1975fg}. 

At high energy (or high temperature) the barrier can be crossed classically by sphalerons  \cite{Klinkhamer:1984di,McLerran:1990de,Moore:2010jd}. 
Within the Standard Model such transitions violate the baryon number conservation and are believed to have occurred in the hot electroweak plasma in the expanding early Universe \cite{tHooft:1976rip,Rubakov:1996vz} satisfying two of Sakharov's conditions for baryogenesis \cite{Sakharov:1967dj}. Unfortunately, the temperature of the electroweak phase transition is too high ($T_\text{EW}\approx 160\ \si{\GeV}$) to observe these processes in the laboratory. 

However, the temperature of the QCD phase transition is 3 orders of magnitudes smaller and the QCD plasma is created in heavy-ion collisions at the Relativistic Heavy Ion Collider (RHIC) and the Large Hadron Collider (LHC). In QCD, sphaleron transitions are accompanied by a chirality flip of the light quarks, representing a ``local $P$- and $CP$-violation" \cite{Kharzeev:1998kz,Kharzeev:2020jxw}. 

The chiral magnetic effect (CME) opens the possibility to detect the topological transitions by converting the chirality asymmetry generated by them into a fluctuation of an electric current, in an external magnetic field \cite{Kharzeev:2004ey,Kharzeev:2007jp,Fukushima:2008xe}. 
In heavy-ion collisions, the produced quark-gluon plasma is penetrated by a strong magnetic field created mostly by the spectator protons \cite{Kharzeev:2007jp}. Therefore, the chirality imbalance generated by topological transitions induces an electric charge separation along the direction of magnetic field, i.e. relative to the reaction plane of a heavy-ion collision.

The experimental observable directly linked to fluctuations of electric current was proposed in~\cite{Voloshin:2010ut} (see \cite{Kharzeev:2015znc} for a review and discussion):   \begin{equation}
 \cos(\Delta\phi_\alpha + \Delta\phi_\beta) \propto 
  \frac{\alpha\beta}{N_\alpha N_\beta} \bigl(
  <J_\perp^2> - <J_\parallel^2> \bigr) \,,
\label{eq:cosJ2}
\end{equation}
where $\Delta \phi$ is the angle relative to the reaction plane of the produced hadrons, with $\alpha,\beta = \pm 1$ indicating positively and negatively charged hadrons. As shown in (\ref{eq:cosJ2}), this experimental observable can be related \cite{Fukushima:2009ft} to the fluctuations in the current $J_\perp$ perpendicular to the magnetic field $B$ and the current $J_\|$ along $B$, where $N_+$ and $N_-$ is the number of positive and negative hadrons, respectively.

To evaluate the observables (\ref{eq:cosJ2}), one needs to use a model for generating the chirality imbalance, and then include the CME and the corresponding hydrodynamical modes (e.g., the chiral magnetic wave \cite{Kharzeev:2010gd}) in the description of the expanding quark-gluon plasma, see \cite{Hirono:2014oda,Shi:2017cpu,Shi:2018sah,Inghirami:2019mkc,Ghosh:2021naw,Cartwright:2021maz,Kharzeev:2022hqz} for different implementations of this program. At present, the major source of uncertainty in the model calculations is the spatial and temporal distribution of the chirality imbalance. While the sphaleron rate in a magnetized strongly coupled plasma has been evaluated in holography \cite{Basar:2012gh}, the resulting spatial and temporal correlations of the electric currents that determine the value of (\ref{eq:cosJ2}) have not been computed.

The goal of the present paper is to fill this gap, and use a microscopic field theoretical model to describe topological transitions in a hot non-Abelian plasma and the resulting correlations  of electric currents. In particular, we will be interested in the spatial correlations of the electric currents, as the size of these correlations strongly affects the observables (\ref{eq:cosJ2}). 

The observed hydrodynamical behavior of the quark-gluon plasma suggests that it is strongly coupled, see e.g.~\cite{Shuryak:2004cy,Shuryak:2008eq}. 
In the context of strongly coupled quantum field theories, holography provides a valuable toolkit to address questions about real-time dynamics.
Within holography, the real-time dynamics of the CME~\cite{Yee:2009vw,Gynther:2010ed,Landsteiner:2013aba} was first studied in \cite{Lin:2013sga,Ammon:2016fru} which was more recently extended to the back-reacted case \cite{Ghosh:2021naw,Grieninger:2021zik,Cartwright:2021maz, Grieninger:2023myf}.

\section{Holographic Setup}

In the presence of dynamical gauge fields
axial current is no longer conserved
\begin{equation*}
   \partial_\mu J_{5}^\mu =c_\text{strong}\, \text{tr}G\wedge G
   +c_\text{em}\!\left(3 F \wedge F+F^{(5)}\wedge F^{(5)}\right).
\end{equation*}

The necessary gravity degrees of freedom to incorporate the non-Abelian chiral anomaly in holography are the following: the work of Klebanov, Ouyang, and Witten \cite{Klebanov:2002gr} argued that the anomaly emerges  from the various form fields on the cycles in the internal part of the 10D background. In the case of the $\mathcal N = 1$ cascading $SU(N + M) \times SU (N)$ gauge theory, it arises from the two-form $F_3 = \dd C_2$ on the three-cycle in the $T^{1,1}$ geometry which renders the bulk vector field massive. Following this philosophy, \cite{Gursoy:2014ela} wrote down a top-down inspired holographic model including the dilaton. The dilaton degrees of freedom are not important for the  discussion of chiral anomaly since it does not depend on the metric (see e.g. \cite{Anastasopoulos:2006cz}). Therefore we freeze the expectation value of the dilaton to a nonsingular value, thus fixing the mass of the axial gauge field and work with the minimal bottom up model of \cite{Jimenez-Alba:2014iia}.

We consider the holographic $U(1)_A\times U(1)_V$ St\"uckelberg model established in \cite{Jimenez-Alba:2014iia,Jimenez-Alba:2015awa}
\begin{widetext}
\begin{align}\label{eq:action}
S= \ \frac{1}{2\kappa_5^2}\int_{\mathcal{M}}\dd^5x\sqrt{-g}&\left[R+\frac{12}{L^2}-\frac{1}{4}F^2-\frac{1}{4}F_{(5)}^2 +\frac{m_s^2}{2}(A_m-\partial_m\theta)^2 \right.\nonumber\\
&\left.\quad+\frac{\alpha}{3} \epsilon^{mnklp} (A_m-\partial_m\theta)\left( 3 F_{nk}F_{lp}+F^{(5)}_{nk}F^{(5)}_{lp}\right) \right]+S_{bdy}+S_{ct}
\end{align}\end{widetext}
with the axial field strength $F_{(5)}=\mathrm{d}A$, the vector field strength $F=\mathrm{d}V$ and the St\"uckelberg (pseudo)scalar $\theta$ which renders the axial gauge field massive while preserving gauge invariance. The strength of the Abelian $U(1)^3_A$ and $U(1)_A\times U(1)_V^2$ anomaly is governed by the parameter $\alpha$ in front of the mixed Chern-Simons term that couples the axial and vector gauge fields. Similarly, the strength of the non-Abelian anomaly is governed by the parameter $m_s$ that determines the mass of the axial gauge field and thus its anomalous dimension. Note that both couplings $\alpha$ and $m_s$ may be separately tuned to different values which we will utilize in the manuscript. 

In this work, we will focus on the explicit breaking of the $U(1)_A$ due to the chiral anomaly and hence consider massless fermions for clarity. Moreover, we aim to study correlations of the electric current due to topological fluctuations which generate axial charge dynamically. Thus, we start with a background geometry that does not contain any finite charge densities but a (Abelian) magnetic field in the $z$-direction
\begin{align}
   & V_\mu=(0,0,- B/2\,y,B/2\,x,0),\ A_\mu=\bm{0},\ \theta=0\label{eq:back1} .
\end{align}
The corresponding background metric is given by the magnetic brane \cite{DHoker:2009mmn} in infalling Eddington-Finkelstein coordinates
\begin{align}
    \mathrm{d}s^2&=\frac{1}{u^2}\left(-f(u)\dd t^2-2\dd t\dd u+v(u)^2\dd x^2\right.\nonumber\\&\left.+v(u)^2\dd y^2+w(u)^2\dd z^2\right).\label{eq:back2}
\end{align}
which takes the anisotropy caused by the magnetic field at equilibrium into account. Unfortunately, the background geometry is not known analytically for all values of the magnetic field and we have to construct it numerically.

The equations of motion associated with the action eq.~\eqref{eq:action} are given by
\begin{align}
\label{eq:eoms}
    &m_s^2\nabla_{m}(A^m-\partial^m\theta)=0\\
    &  \nabla_n F^{nm}+2\alpha \epsilon^{mnklp} F_{nk}F^{(5)}_{lp}=0\,,\\
    & \nabla_m F_{(5)}^{mn}-m_s^2(A^n -\partial^n\theta)\nonumber\\&+\alpha  \epsilon^{mnklp} \left( F_{nk}F_{lp}+F_{nk}^{(5)}F_{lp}^{(5)} \right)=0\,\label{eq:vector}, \\
    &  R_{mn}\!-\!\frac R2 g_{mn}\!-\!6 g_{mn}\!-\!\frac{1}{2} F_{mk}F_{n}^{\ p}
 +\frac{1}{8} (F^2+F_{(5)}^{2}) g_{mn} \nonumber\\&-\frac{1}{2} F^{(5)}_{mk}F_{n}^{(5) p }-\dfrac{m_s^2}{2}(A_m - \partial_{m}\theta)(A_n - \partial_{n}\theta)\nonumber\\&+\frac{1}{4}m_s^2 (A_k-\partial_{k}\theta)(A^k-\partial^{k}\theta)g_{mn}=0\,.
\end{align} 
Note that we do not recover the decoupled axion as wave equation upon sending $A_\mu$ and $V_\mu$ to zero since eq.~\eqref{eq:vector} would impose the constraint $\dd\theta=0$.
We consider linearized spacetime dependent fluctuations about the background specified in eq.~\eqref{eq:back1} and eq.~\eqref{eq:back2} by considering
\begin{align}
    V_\mu&=\,(0,0,- B/2\,y,B/2\,x,0)\nonumber\\&+\varepsilon e^{i(-\omega t+\bm{k}\cdot\bm{x})}\{v_t(u),0,v_x(u),v_y(u),v_z(u)\},\\ A_\mu&=\varepsilon e^{i(-\omega t+\bm{k}\cdot\bm{x})}\{a_t(u),0,a_x(u),a_y(u),a_z(u)\},\\ \theta&=\varepsilon e^{i(-\omega t+\bm{k}\cdot\bm{x})}\vartheta(u)
\end{align}
to linear order in $\varepsilon$, where $\omega$ and $\bm{k}\equiv(k_x, k_y, k_z)$ are Fourier frequency and momentum, respectively. Owing to the mass of the gauge field the coefficient of the leading and subleading modes in the asymptotic expansion are modified to
\begin{align}
  &  a_\mu(u)\sim \,a_\mu^{(l)}\,u^{-\Delta}(1+..)+\partial_\mu \vartheta^{(l)}+a_\mu^{(sl)}\,u^{2+\Delta}(1+..),\nonumber\\&  \Delta\equiv -1 + \sqrt{1+   m_s^2}.
  \end{align}
  The superscripts (l) and (sl) refer to leading and subleading, respectively, and where fractional exponents proportional to multiples of $\Delta$ and $2+\Delta$ also appear in the expansion of the vector gauge field and the St\"uckelberg scalar. 

The numerical simulations are conducted for parameters relevant for the Quark-Gluon plasma (QGP) by matching the entropy density and Abelian anomaly to three flavor QCD results as explained in \cite{Ghosh:2021naw,Grieninger:2021zik,Grieninger:2021rxd}.
The coefficient of the mixed Chern-Simons term   is denoted as $\mathcal{A}_{CS} = \frac{\alpha}{2 \kappa_5^2}$. To estimate $\kappa_5$, we match the entropy of a black brane $s_{BH} = \frac{4\pi^4T^3}{2 \kappa_5^2}$ to the Stefan-Boltzmann value of entropy density in three flavor QCD, including the strange quark.
The Stefan-Boltzmann value of entropy density $s_{SB} = 4\left(\nu_b + \frac{7}{4}\nu_f\right)\frac{\pi^2T^3}{90}$ is used, but a reduction factor of $3/4$ is taken into account due to moderate temperatures (around $T=300\si{\MeV}$) which is known in the context of $\mathcal N=4$ super-Yang Mills theory~\cite{Gubser:1996de}.\footnote{Lattice simulations indicate a relative factor of 0.8 (see for example \cite{Borsanyi:2013bia}).} Matching the model to QCD gives $\kappa_5^2 = \frac{24\pi^2}{19} \approx 12.5$.
The axial anomaly in three flavor QCD is $\mathcal{A}_{QCD} = \frac{1}{8\pi^2}$. By matching $\mathcal{A}_{CS}$ to $\mathcal{A}_{QCD}$, the Chern-Simons coupling is determined as $\alpha = \frac{6}{19} \approx 0.316$.

All numerical simulations are performed using pseudospectral methods~\cite{Boyd1989ChebyshevAF}.\footnote{See appendix A of \cite{Grieninger:2020wsb} for an introduction applied to a related setup.}

\section{Axial charge relaxation rate}
\begin{figure}[h!]
    \centering
\includegraphics[width=0.99\linewidth]{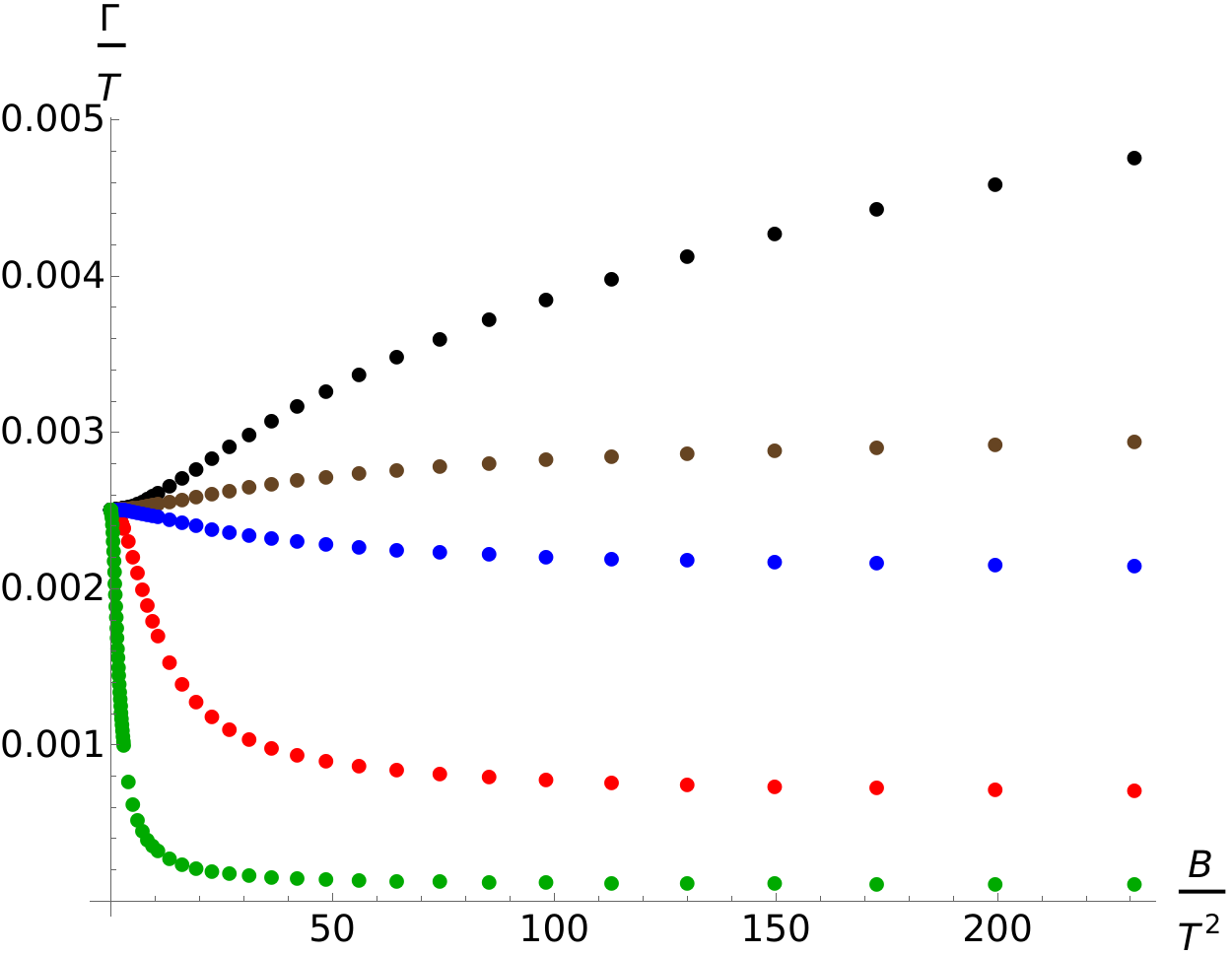}
    \caption{Axial charge relaxation rate as a function of $B/T^2$ for five different values of the Abelian anomaly ($\alpha=\{0,0.1,0.15,0.32,2\}$ corresponding to (black, brown, blue red and green)). We fixed $m_sL=0.04$. For small magnetic fields the dependence on $B/T^2$ is quadratic.}
    \label{fig:0}
\end{figure}

Owing to the explicit breaking of the axial $U(1)$ symmetry, the chiral-magnetic wave (CMW) \cite{Kharzeev:2010gd} is no longer gapless at zero wave vector as was observed in the probe approximation in \cite{Jimenez-Alba:2014iia} (for completeness we display the dispersion relation for the full system in fig~\ref{fig:gappedCMW} of appendix \ref{app:cmw}). For small momenta (along the magnetic field) the CMW is purely diffusive and only starts propagating for $k_\|>k_{\|,c}$. If the symmetry breaking is small enough, the dynamics can still be treated within quasihydrodynamics as was explicitly demonstrated in a similar setup in \cite{Ammon:2021pyz}. In this section, we focus on the $k\equiv0$ dynamics which determines the rate of axial charge relaxation $\Gamma$ according to $n_5(t)\sim e^{-\Gamma\, t}\,n_5(0)$ \cite{Jimenez-Alba:2014iia,Jimenez-Alba:2015awa,Grieninger:2023myf}. \footnote{In particular in \cite{Grieninger:2023myf}, we explicitly demonstrate that the homogeneous real-time dynamics decays exponentially with $\Gamma$ obtained from the quasinormal modes} In holography, dispersion relations are encoded in the so-called quasinormal modes (QNMs) which are solutions to the linearized fluctuation equations in the absence of sources. The QNMs in this model at zero $m_s$ describing the CME and axial charge equilibration for the unbroken $U(1)_A\times U(1)_V$ symmetry were computed in~\cite{Ammon:2016fru,Grieninger:2016xue}. In fig \ref{fig:0}, we show the axial charge relaxation rate as a function of the magnetic field at fixed $m_s$ and for five different values of the Abelian anomaly ($\alpha=\{0,0.1,0.15,0.32,2\}$ corresponding to (black, brown, blue, red and green)). For small magnetic fields, the dimensionless axial charge relaxation rate $\Gamma/T$ grows (decreases) quadratically as a function of $B/T^2$ for $\alpha=\{0,0.1\}$ ($\alpha=\{0.15,0.32,2\}$). As was also observed in real-time simulations \cite{Grieninger:2023myf}, large magnetic fields and sufficiently large strength of the Abelian anomaly protect axial charge and lower the axial charge relaxation rate. At large magnetic field the dimensionless axial charge relaxation rate decreases linearly in $B/T^2$ for $\alpha=\{6/19,2\}$ where the larger value of $\alpha$ leads to a smaller $\Gamma/T$ at large $B/T^2$. Increasing the strength of the non-Abelian anomaly by increasing $m_s$ increases $\Gamma$, while increasing the strength of the Abelian anomaly $\alpha$ (at finite $B$) decreases $\Gamma$ as was shown in \cite{Grieninger:2023myf}.
\section{Chern-Simons diffusion rate}\label{sec:CSdiff}
The rate at which the Chern-Simons number changes is known as the Chern-Simons diffusion rate, denoted by $\Gamma_\text{CS}$. This rate represents the probability of a Chern-Simons number changing process to occur per unit volume and per unit time interval.

The Chern-Simons rate is given by the symmetrized Wightman
correlator of topological charge in the zero frequency and momentum limit. In holography, the symmetrized Wightman function is related to the retarded Green's function using the fluctuation-dissipation theorem (assuming detailed balance)~\cite{Son:2002sd}. The Chern-Simons diffusion rate in a magnetic field (without any other matter fluctuations and anomalies present) was calculated in \cite{Basar:2012gh}.\footnote{The Chern-Simons diffusion dynamics was also considered in \cite{Gursoy:2012bt,Bigazzi:2018ulg,Craps:2012hd,Bu:2014cca,Jahnke:2014sla,Son:2002sd,Iatrakis:2015fma,Iatrakis:2014dka}.} In the setup we are considering there is no axial charge left at late times ($\omega\to 0$ limit). If axial charge relaxes sufficiently slowly, we can still describe the weakly explicitly broken symmetry within the framework of quasihydrodynamics.
\begin{figure}[h!]
    \centering
\includegraphics[width=0.99\linewidth]{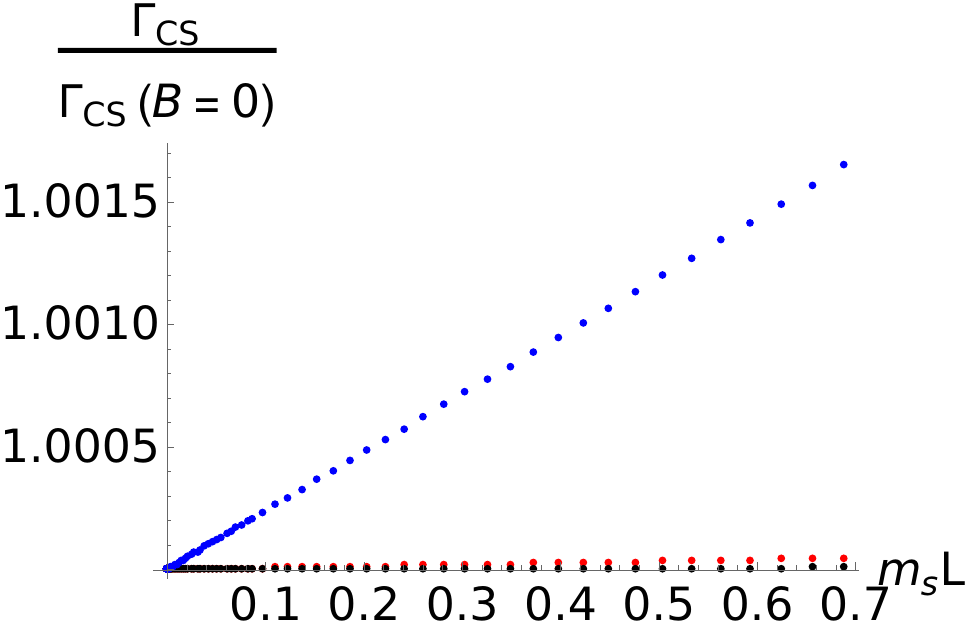}
    \caption{Chern-Simons diffusion rate for $B\approx 0.22 T^2$ normalized to the Chern-Simons rate at $B=0$ for $\alpha=2$ (blue), $\alpha=6/19$ (red) and $\alpha=0$ (black).}
    \label{fig:CSdiff1}
\end{figure}
The axial charge relaxation rate is then related to the Chern-Simons diffusion rate by \cite{Bigazzi:2018ulg}
\begin{equation}
     \frac{\dd n_5}{\dd t}=-2q=-\frac{2\Gamma_\text{CS}}{\chi_5\,T}n_5=-\Gamma \,n_5,
\end{equation}
where $q$ is the topological charge density and $\chi_5$ the axial charge susceptibility. Since $\chi_5$ has an anomalous dimension due to the anomalous dimension of the axial current, $\Gamma_\text{CS}$ also inherits this anomalous dimension.
In figure \ref{fig:CSdiff1}, we show the Chern-Simons diffusion rate at finite magnetic field normalized to the Chern-Simons diffusion rate at zero magnetic field as a function of the strength of the non-Abelian anomaly for three different values of the Abelian anomaly ($\alpha=0$, black; $\alpha=6/19$, red; $\alpha=2$, blue). In all three cases, the Chern-Simons diffusion rate increases even though the effect is smaller for smaller $\alpha$. It is evident, that the Chern-Simons rate tends linearly to zero for $m_sL\to 0$. Note that in our macroscopic model \eqref{eq:action} the fermion current is always coupled to the gluons and we did not introduce any mechanism to decouple them from one another. This implies that since the fermion current interacts with the gluons, fluctuations of Chern-Simons number will induce the non-Abelian anomaly. In this model, setting the non-Abelian anomaly to zero implies then that we remove fluctuations of Chern-Simons number. In order to connect our results to \cite{Basar:2012gh} which computed the Chern-Simons rate in strong magnetic fields, we studied the scaling of the Chern-Simons rate for small and large magnetic fields. As shown in left side of figure \ref{fig:CSapp2} in appendix \ref{app:CSdiff}, $\Gamma_\text{CS}$ increases quadratically for small $B/T^2$ and linearly at large $B/T$ in agreement with~\cite{Basar:2012gh}. On the right side of figure \ref{fig:CSapp2}, we show that the Abelian anomaly $\alpha$ increases the Chern-Simons diffusion rate, an effect that becomes more pronounced at larger magnetic fields.

\section{Correlations of electric current}
In this section, we study the correlations of electric current fluctuations.
Our procedure is the following: for a given value of the magnetic field, we numerically construct the background geometry given by the magnetic black brane. On top of this background we compute the retarded two-point correlators of the electric current (along the magnetic field) as a function of the three momentum $\bm{k}$ at finite $\omega$.
To isolate the topological dynamics from other correlations such as thermal fluctuations and to subtract any contact terms, large $\bm{k}$ divergences and finite $\omega$ contributions, we define the following subtracted correlator 
    \begin{equation}\label{eq:subcorr}
        \Delta G^\text{ret}_{J^zJ^z}(\omega,\bm{k})\equiv G^\text{ret}_{J^zJ^z}(\omega,\bm{k},m_s)-G^\text{ret}_{J^zJ^z}(\omega,\bm{k},0),
    \end{equation}
which is the difference of the two-point function with topological fluctuations and the two-point function with topological dynamics switched off (only Abelian anomaly). 
Considering this observable has several advantages from a physical and technical point of view. The subtracted correlator isolates the topological dynamics since we subtract any other contributions. The CME current $\langle J^z\rangle$ has conformal dimension three independent of the value of $m_s$. From a technical point of view we note that the asymptotic expansions of the fluctuations have logarithmic divergences proportional to the source, momentum, frequency and magnetic field. These divergences have to be removed by the appropriate counter terms which break conformal invariance and thus inherently cause scale dependent contact terms (see \cite{Horowitz:2008bn,Grieninger:2022yps} for a discussion). However, since these divergences are independent of $m_s$ we automatically remove any scheme dependent contact terms by considering \eqref{eq:subcorr} since they cancel exactly. Finally, it's advantageous to consider the subtracted correlator \eqref{eq:subcorr} over the two-point function $G^\text{ret}_{J^zJ^z}(\omega,\bm{k},m_s)$ for the following reason: the two-point functions generally diverge for large momentum and frequency. The large momentum frequency could be subtracted by subtracting the zero temperature result.\footnote{Instead of the zero temperature result one could also subtract the result at a different temperature as done in \cite{Gursoy:2012bt}. In our setup, this is not possible since changing the temperature means also changing the magnetic field.} However, since we are working at finite frequency the two-point function would still not go to zero for large momenta (as we prefer for performing the inverse Fourier transform). Considering \eqref{eq:subcorr} removes this unwanted correlation elegantly. We consider the absolute value of the retarded two-point function in position space which is motivated by the standard definition
of correlation radius from a form factor, where it is defined through a modulus squared of this quantity.

After computing the subtracted correlator \eqref{eq:subcorr} numerically, we translate the result into real (position) space by performing an inverse (discrete) Fourier transform. In order to characterize the range of correlations, we compute the root mean square defined by
    \begin{equation}\label{eq:rms}
        x_\text{rms}=\sqrt{\frac{\int \dd x\, x^2\, |\Delta G^\text{ret}_{J^zJ^z}(x)|}{\int \dd x\,|\Delta G^\text{ret}_{J^zJ^z}(x)|}}.
    \end{equation}
     At finite $\omega$ the two-point functions are generally complex and we consider the absolute value. 
    Since we subtract all other correlations, the fluctuations of electric current are driven by the axial charge induced by the topological transition and the spatial profile of the electric current two-point function encodes the information about spatial profile of induced axial charge for a given magnetic field and time interval.
\subsection{Spatial distribution}
\begin{figure*}
    \centering
\includegraphics[width=0.48\linewidth]{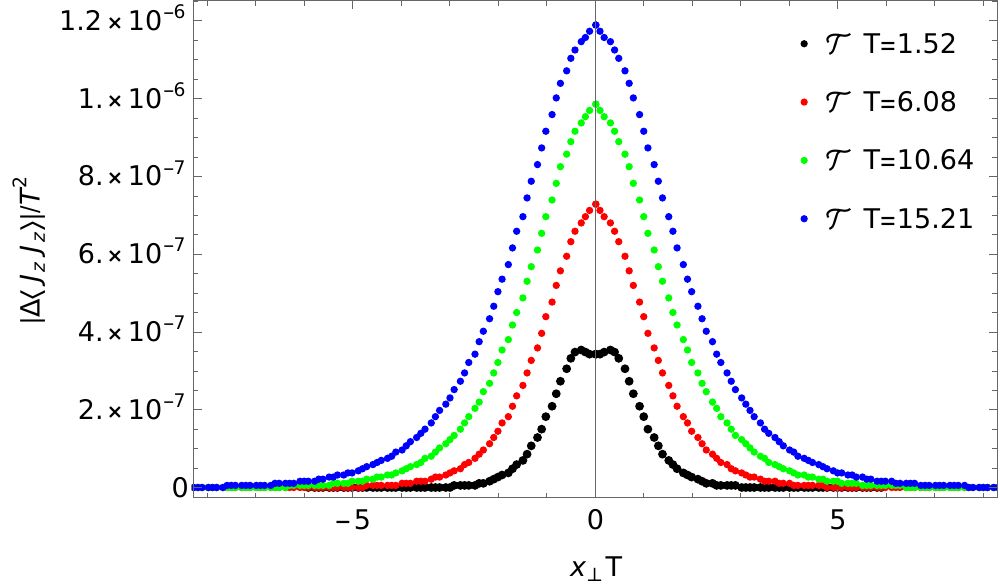}
\includegraphics[width=0.48\linewidth]{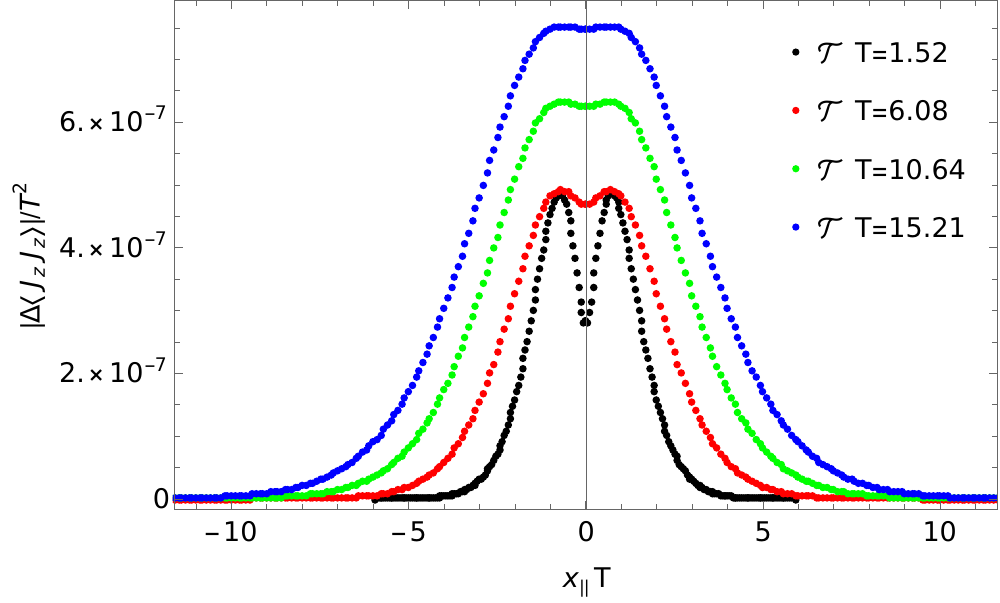}
\includegraphics[width=0.48\linewidth]{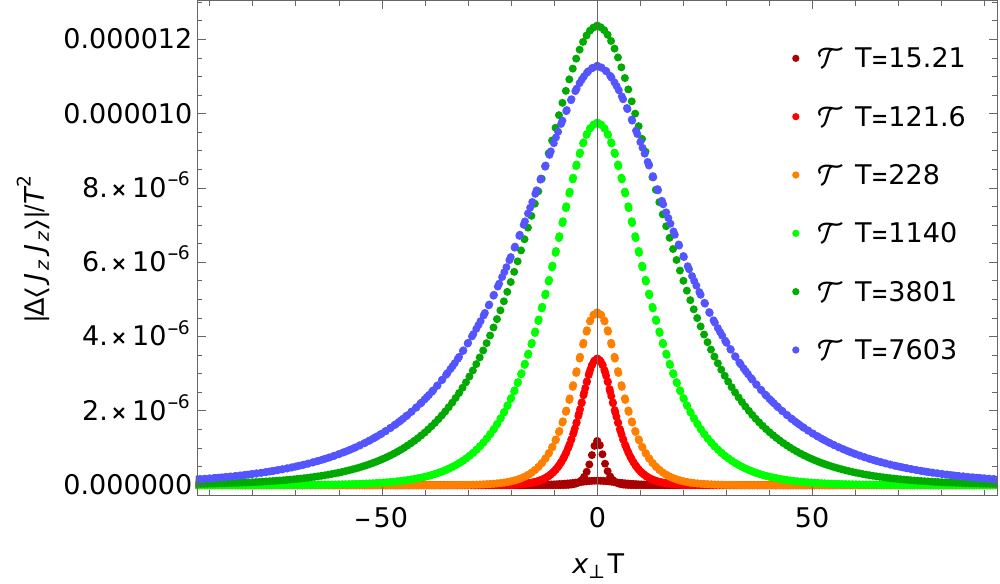}
\includegraphics[width=0.48\linewidth]{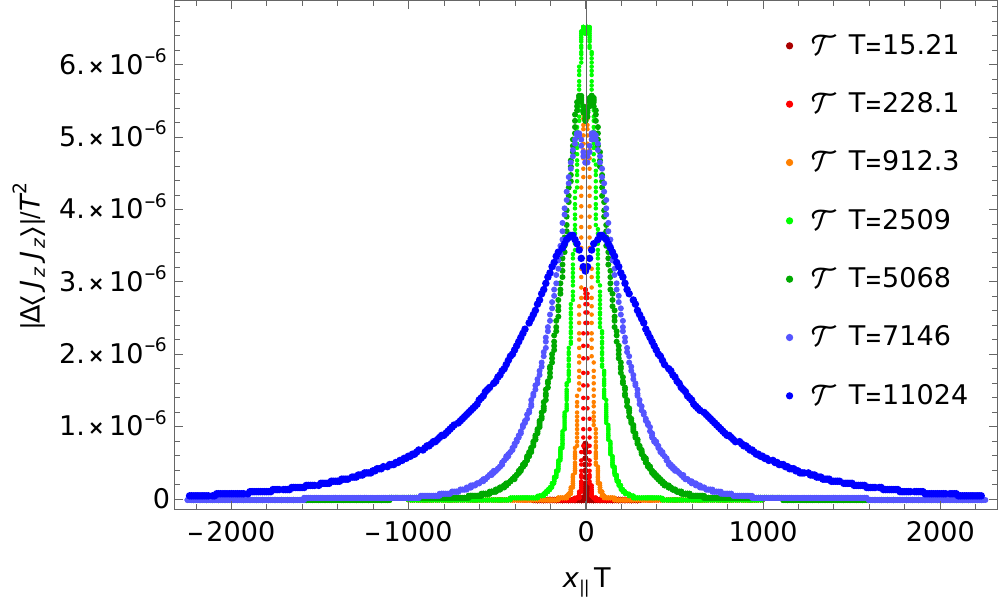}
    \caption{Initial spatial distribution (top panel) and late time spatial distribution (lower panel) of the subtracted electric current two-point function defined in eq. \eqref{eq:subcorr}. We fixed $B\approx 0.22 T^2$, $\alpha=6/19$ and $m_sL=0.04$. The plots are shown in units of temperature.}
    \label{fig:01}
\end{figure*}
To illustrate the real-time dynamics, figure \ref{fig:01} depicts the spatial profile of the subtracted electric current two-point function \eqref{eq:subcorr} at different moments in time. The top panel captures the buildup of the correlations at early times and the lower panel depicts the late time dynamics. Transverse and longitudinal refer to the direction of the momentum (Fourier space) or Cartesian coordinate with respect to the direction of the magnetic field. For the earliest shown time interval (black curve, top panel) the transverse (left) and longitudinal distribution (right) show two peaks. This might be the strong coupling analog of the two chiral fermions in the weak coupling picture. Increasing the length of the time interval the distributions increase in spatial extent and magnitude. From the top panel it is evident that the area between the two off axis peaks fills up corresponding to filling up the sphaleron shell. After reaching the axial charge relaxation time $\tau_{5,\text{rel}}=2501/T$, the magnitude of the distributions starts to decrease while their spatial extent continues to increase (lower panel). Furthermore, the two peaks start appearing again in the longitudinal distribution (lower right panel).

\subsection{Spatial extent of the correlations}
\begin{figure*}
    \centering
\includegraphics[width=0.48\linewidth]{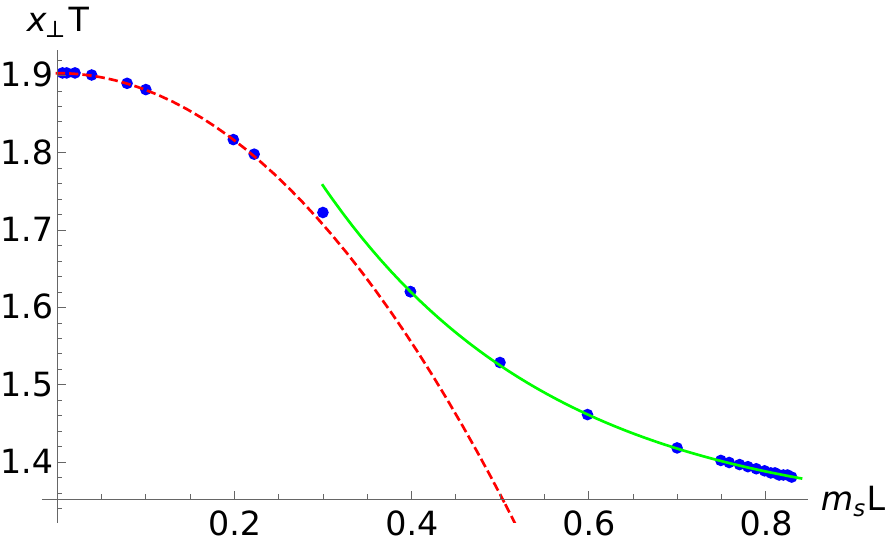}
\includegraphics[width=0.48\linewidth]{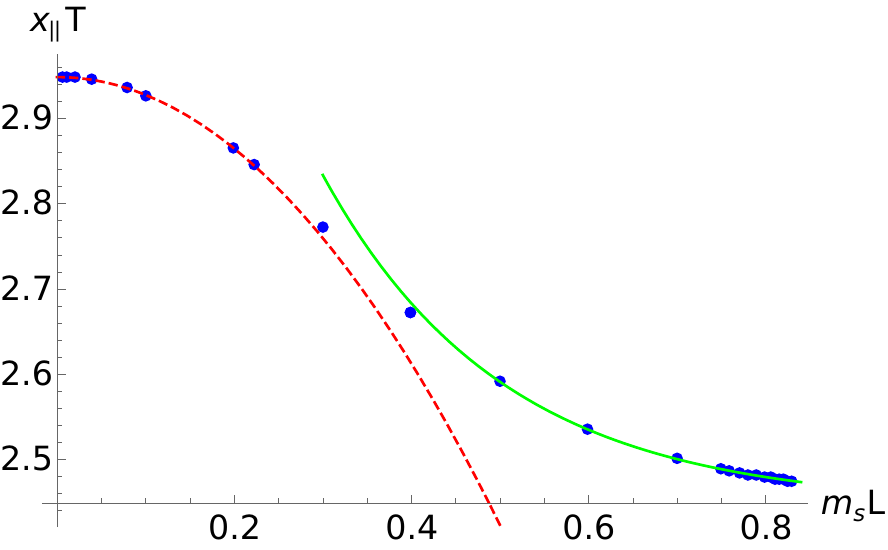}
    \caption{Root mean square $x_{\text{rms},\{\perp,\|\}}$ \eqref{eq:rms} as a function of the strength of the non-Abelian anomaly $m_sL$. We fixed $B\approx 0.22 T^2, T{\cal T}=15.21$ and $\alpha=6/19$.}
    \label{fig:1}
\end{figure*}

Before we give an estimate for the spatial extent of the electric correlations, we investigate the dependence on the strength of the non-Abelian anomaly $m_s L$ which governs the anomalous dimension of the axial current. 
At small $m_sL$, the coupling dependence is well fitted by \begin{align}
    &x_\perp T=1.90- 2.18 (m_sL)^{2.00}\\
    &x_\perp T=2.95- 2.10 (m_sL)^{2.00}
\end{align}
as indicated by the red dashed in fig \ref{fig:1}. It makes sense that the size decreases for increasing $m_s$ since we observed in figure \ref{fig:CSdiff1} that the Chern-Simons diffusion rate increases with $m_s$ making topological transitions more likely. At large couplings the data are best fitted by an exponential (green dashed line in fig \ref{fig:1}) and we find $x_\perp T=1.32+1.38 e^{-3.86 (m_sL)}$ as well as $x_\| T=2.45+1.69 e^{-4.92 (m_sL)}$. If we extrapolate this fit to $m_sL\to\sqrt{3}$ (which corresponds to the largest anomalous dimension allowed for the current) and compare the size to the fit at small coupling extrapolated to $m_sL=0$, we find
\begin{equation}
    \frac{x_\|(m_sL=\sqrt{3})}{x_\|(m_sL=0)}=0.83,\quad \frac{x_\perp(m_sL=\sqrt{3})}{x_\perp(m_sL=0)}=0.70.\label{eq:sizevariation}
\end{equation}
However, for the specific value of $B/T^2=0.22$ that we chose for most simulations, the axial charge relaxation rate and thus $m_s L$  should be very small otherwise the dynamics could not be captured successfully within hydrodynamics (as seems to work very well in the case of heavy-ion collisions) and the variation of the size should be much smaller than the upper limit we estimated in eq.~\eqref{eq:sizevariation}. Hence, in the following we fix $m_sL=0.04$.

\begin{figure*}
    \centering
\includegraphics[width=0.48\linewidth]{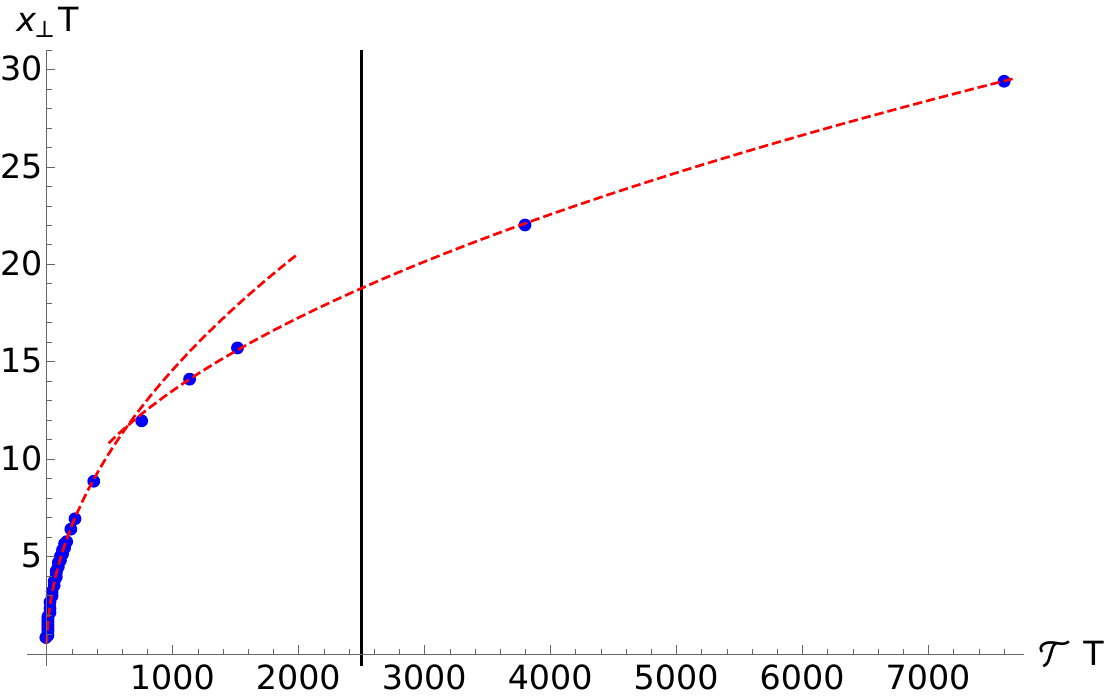}
\includegraphics[width=0.48\linewidth]{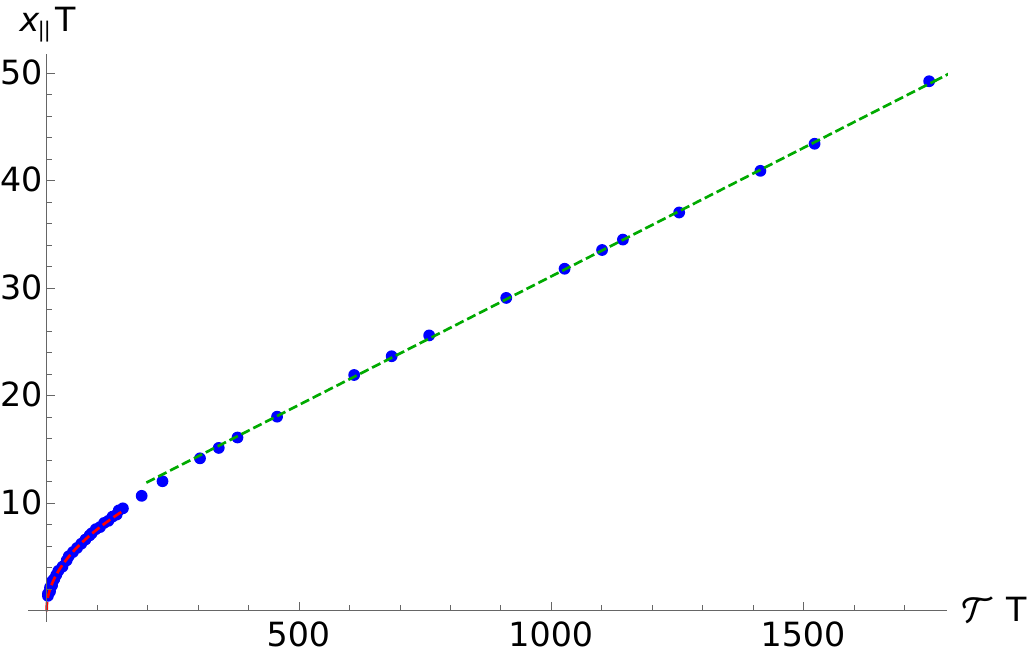}
    \caption{Transverse (left) and longitudinal (right) size characterized by the root mean square $x_{\text{rms},\{\perp,\|\}}$ \eqref{eq:rms} as function of the time interval ${\cal T}$. We fixed $B\approx 0.22 T^2$, $\alpha=6/19$ and $m_sL=0.04$. }
    \label{fig:2}
\end{figure*}

In the left panel of figure \ref{fig:2}, we depict the growth of the root mean squared of the transverse distribution as a function of time at fixed $B,\alpha$ and $m_s$. The axial charge relaxation time is indicated by the vertical black line. There are two different regimes in which the growth of the root mean square is diffusive with different growth rate
\begin{align}
    & x_\perp T=0.12 + 0.46\,  \sqrt{{\cal T} T},&\text{for } {\cal T} T \text{ small}\\
   & x_\perp T=4.46 + 0.29\,  \sqrt{{\cal T} T},& \text{for } {\cal T} T \text{ large}
\end{align}

The growth of the size in direction along the magnetic field is shown in right panel of \ref{fig:2}. We observe two different scaling regimes which may be fitted by
\begin{align}
    & x_\| T=0.027 + 0.75\,  \sqrt{{\cal T} T},&\text{for } {\cal T} T \text{ small},\\
   & x_\|  T=7.14 + 0.024 \,{\cal T} T,& \text{for } {\cal T} T \text{ large}.
\end{align}
At early times the growth of the longitudinal size is diffusive as in the transverse case. However, at later times we observe ballistic growth which scales linear in the time interval. The real part of the subtracted correlator in longitudinal direction changes sign at very late times, and the real and imaginary part become oscillatory (even though the absolute value is still well behaved); we decided to cut the plot there. 

\subsection{Dependence on magnetic field}
\begin{figure*}
    \centering
\includegraphics[width=0.48\linewidth]{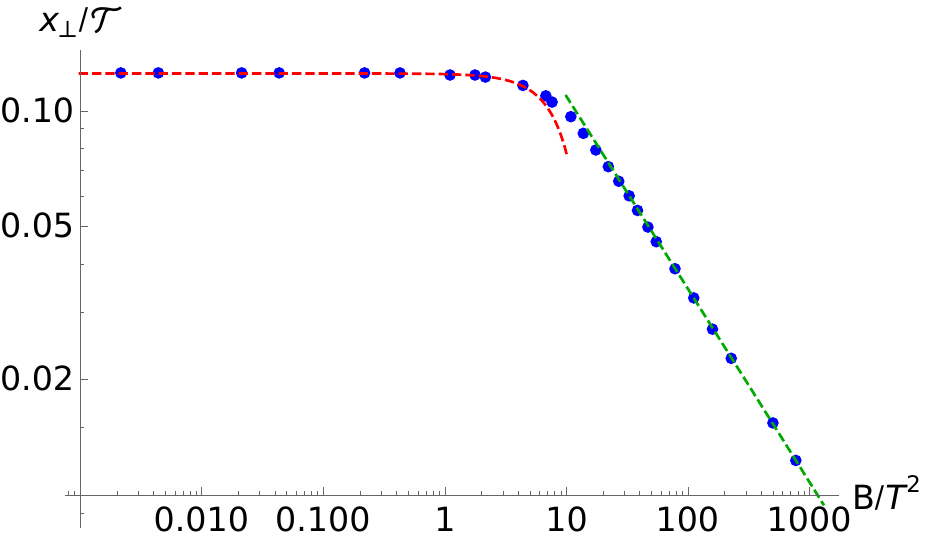}
\includegraphics[width=0.48\linewidth]{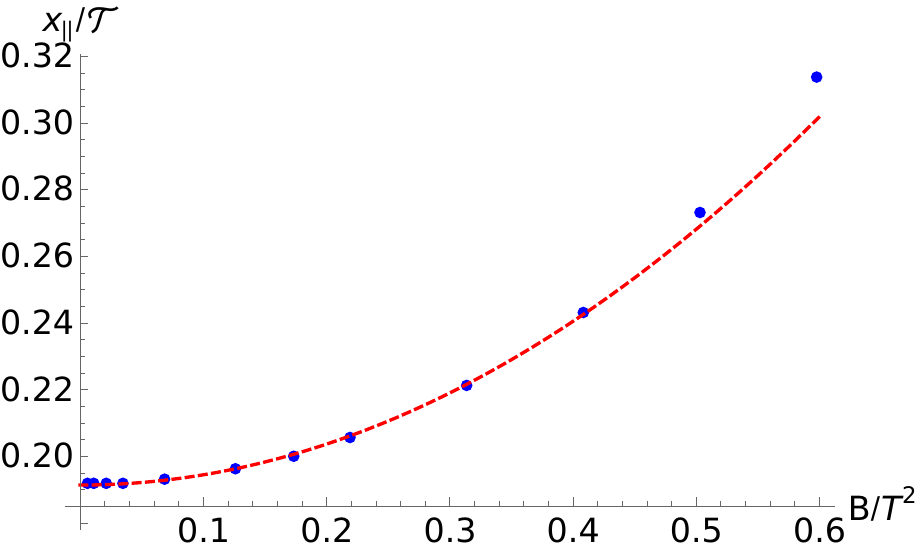}
    \caption{Transverse (left) and longitudinal (right) size as a function of the magnetic field $B$. The size is normalized to the time interval ${\cal T}$. We fixed $T{\cal T}=15.21$, $\alpha=6/19$ and $m_sL=0.04$. The red dashed line and the green dashed line are a fit to the small and large $B/T^2$ behavior, respectively.}
    \label{fig:3}
\end{figure*}

To complete our discussion, we investigate the root mean square as a function of the magnetic field. In figure \ref{fig:3}, we show the magnetic field dependence of the transverse size (left) and longitudinal size (right), respectively. Since there is a one to one correspondence between strength of the magnetic field and temperature, we normalize the size to the length of the time interval instead of the temperature as in previous plots while keeping the dimensionless quantity ${\cal T}T$ fixed. The dependence of the transverse size on the magnetic field is depicted in the right panel of figure \ref{fig:3}. The transverse size decreases with the magnetic field according to 
\begin{align*}
   & x_\perp/{\cal T}=0.12 - 0.00047\,  \left(\frac{B}{T^2}\right)^{2.00},&\text{for } \frac{B}{T^2} \text{ small},\\
   & x_\perp/{\cal T}=-0.00011+ 0.35\left(\frac{T^2}{B}\right)^{0.50},& \text{for } \frac{B}{T^2} \text{ large}.
\end{align*}

The longitudinal size grows with the magnetic field 
\begin{align*}
    & x_\|/{\cal T}=0.19+ 0.31\,  \left(\frac{B}{T^2}\right)^{2.00},&\text{for } \frac{B}{T^2} \text{ small},
    \end{align*}
    as shown in the right panel of figure \ref{fig:3}. 
In the longitudinal case, the peaks become increasingly sharp for larger magnetic fields, making them difficult to reliably resolve numerically. Thus, we decided to not depict the large $B$ behavior. For increasing the magnetic field, the sphaleron becomes more elongated along the magnetic field where its size grows rapidly and the dynamics becomes effectively 1+1 dimensional at large magnetic fields where the physics is determined by the lowest Landau level dynamics. The lowest Landau picture also explains the $1/\sqrt{B}$ scaling of the transverse size. 
    
\section{Conclusions}
In conclusion, we addressed the Chern-Simons diffusion real-time dynamics in strongly coupled field theories subject to strong (external) magnetic field, with a  focus on the interplay of non-Abelian and Abelian anomalies.

We showed that correlations of electric currents are sensitive to correlations of Chern-Simons densities at large distances, which is directly relevant for experimental measurements of the CME~\cite{STAR:2021mii}.
The range of correlations grows with time, exhibiting diffusive behavior perpendicular to the direction of the magnetic field and ballistic behavior parallel to it. This behavior is consistent with sphaleronlike dynamics.

At strong magnetic field, the perpendicular size decreases with the inverse square root of the magnetic field strength (consistent with the lowest Landau level picture). The corresponding longitudinal size grows with the magnetic field. 
This can be viewed as a consequence of the absence of  backscattering (which would introduce chirality nonconservation) in the longitudinal direction at large magnetic fields. The dynamics is confined to the direction parallel to the magnetic field and is effectively 1+1 dimensional.
We checked that the scalings are not dependent on the specific value of $m_sL$ that we used  by performing analogous fits for larger $m_sL$ leading to the same exponents.

In order to give a realistic estimate of the spatial sizes of the electric current correlations, let us express our dimensionless quantities in dimensionful units.
For $T=300\ \si{\MeV},\, B=1\ m_\pi^2,\, {\cal T}= 10\, \text{fm}$ (for $m_sL=0.04$), we estimate the transverse and longitudinal sizes as \begin{equation*}\vspace{-0.18cm}
    x_\perp=1.25\, \text{fm}\qquad\text{and}\qquad x_\|=1.94\,\text{fm}.
\end{equation*}
At zero temperature and without a magnetic field, the average instanton size has been estimated to be $\simeq 0.3$ fm  \cite{Ostrovsky:2002cg,Shuryak:2021iqu,DIAKONOV1988809}. We see that the correlations between electric currents in our case have a significantly larger range. At weak coupling $g$, the size of the sphaleron at temperature $T$ can be estimated as $\sim 1/(g^2 T)$, as the sphaleron is a purely magnetic configuration at the top of the barrier, and its size should be determined by magnetic screening. So sphalerons are large objects (on thermal scale $1/T$) at weak coupling; our study suggests that they are  large at strong coupling as well.

It would be interesting to address the transport effects we investigated in this work within the framework of quasihydrodynamics by extending the hydrodynamic theory of \cite{Ammon:2020rvg} to $U(1)_V\times U(1)_A$ and then breaking $U(1)_A$ explicitly. In the context of explicitly broken $U(1)$ symmetry, it was shown within holography  that such extensions of hydrodynamics are consistent for sufficiently small explicit breaking~\cite{Ammon:2021pyz}.

In realistic heavy-ion collisions, the plasma is rapidly expanding. In~\cite{Grieninger:2023myf}, we studied the homogeneous dynamics of the topological axial charge dynamics. It would be very valuable to extend this discussion to include the spatial dynamics as presented in this work and make the magnetic field time dependent. 

Finally, in our holographic model, we are restricted to sufficiently high temperatures in the plasma phase,  $T \sim 300\si{\MeV}$, since we do not include a realistic behavior of the entropy density near the QCD phase transition. In light of the beam energy scan at RHIC, it would be interesting to investigate the dynamics at lower temperatures and high baryon density by employing holographic models more closely tailored to phenomenology such as V-QCD (see ~\cite{Jarvinen:2022doa} and references therein). Considering these improved holographic models is also interesting (even without magnetic field) in order to compare to lattice QCD results where the sphaleron rate was recently computed in 3 flavor QCD~\cite{Bonanno:2023thi}.

We plan to report on some of these studies in future publications.
 \vskip0.3cm
 
\noindent\textbf{Acknowledgments:} This
work was supported by the Office of Science, Office
of Nuclear Physics, U.S. Department of Energy under Contract No. DE-FG88ER41450.
\appendix
\section{Chiral magnetic wave}\label{app:cmw}
\begin{figure*}
    \centering
    \includegraphics[width=0.49\linewidth]{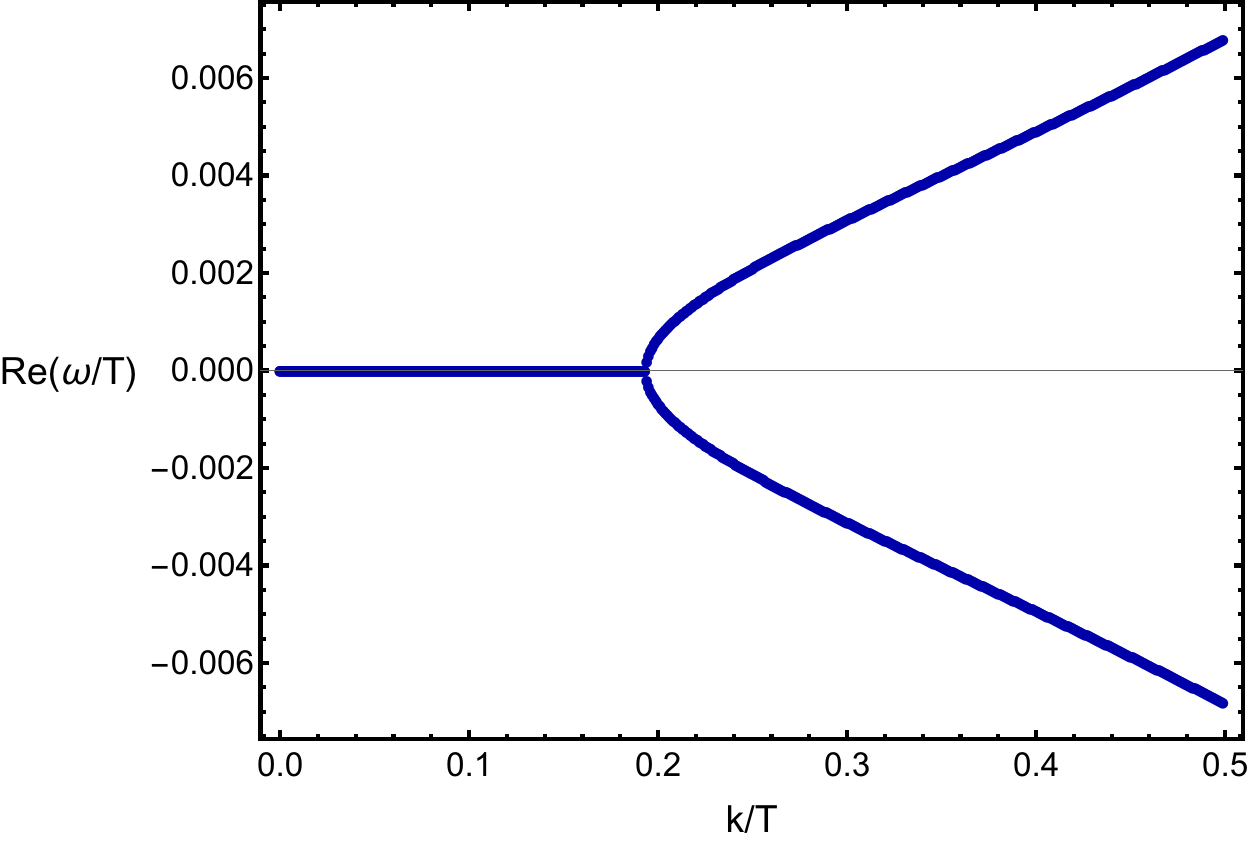} \includegraphics[width=0.49\linewidth]{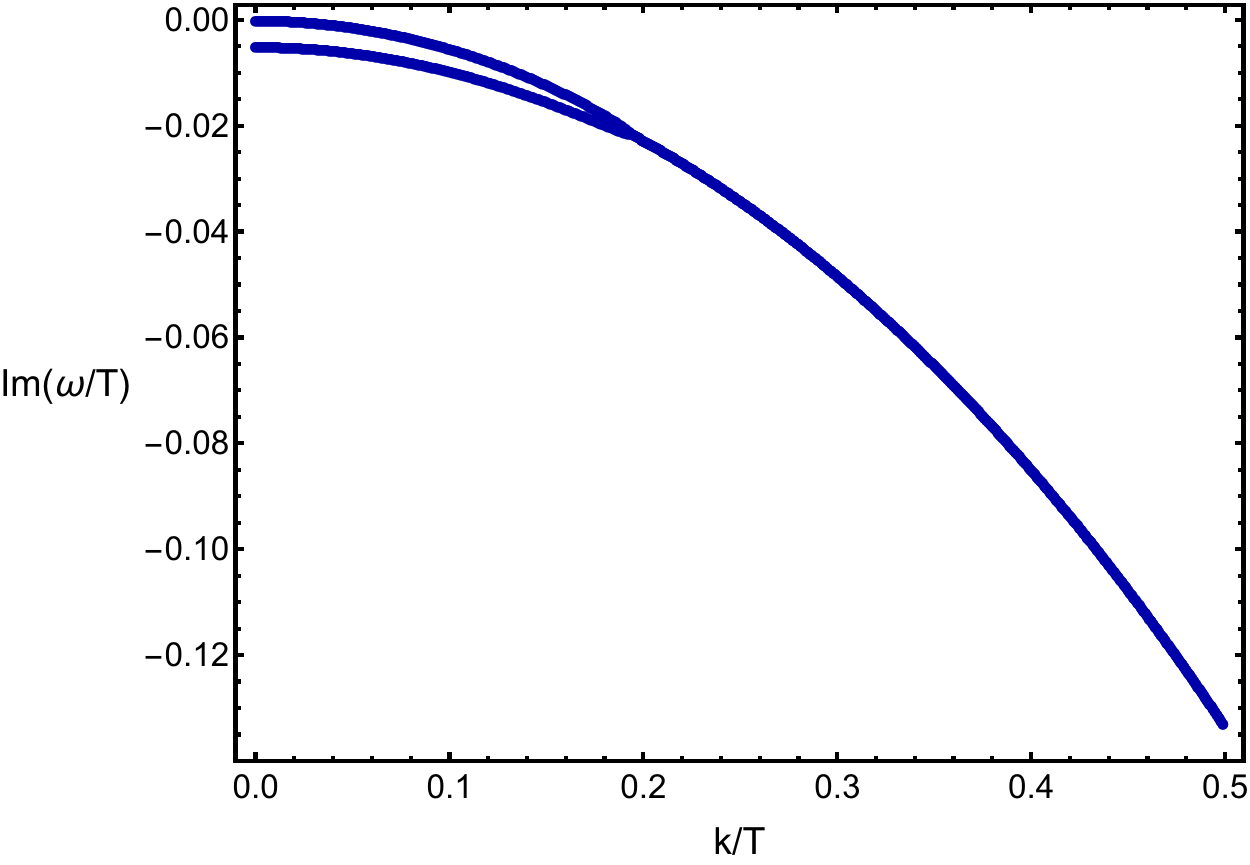}
    \caption{Real (right) and imaginary (left) part of the dispersion relation $\omega(k_\|)$ for finite wave vector (parallel to the magnetic field) at fixed $B/T^2, \alpha=6/19$ and $m_s=0.001$. Finite $m_s$ breaks the $U(1)_A$ explicitly and the chiral magnetic wave s purely diffusive for small wave vectors. \label{fig:gappedCMW}}
    \end{figure*}

For the sake of completeness, we show the momentum dependence of the gapped chiral-magnetic wave (with momentum aligned with the magnetic field) in figure \ref{fig:gappedCMW}.\footnote{For this model in the probe approximation a similar plot was shown in \cite{Jimenez-Alba:2014iia}}
    For small momenta the chiral magnetic wave is nonpropagating and purely diffusive. Above a certain scale, the symmetry is restored and the chiral magnetic wave starts propagating. At zero momentum the imaginary part of the mode corresponding to the unbroken $U(1)_V$ symmetry connects to zero while the imaginary part of the explicitly broken $U(1)_A$ is gapped with the gap we referred to as $\Gamma$ determining the axial charge relaxation time. For a detailed discussion of this behavior in the context of explicitly broken symmetries see for example the review \cite{Baggioli:2019jcm} and references therein.
\section{Chern-Simons diffusion rate and chiral anomaly}\label{app:CSdiff}
\begin{figure*}
    \centering
    \includegraphics[width=0.49\linewidth]{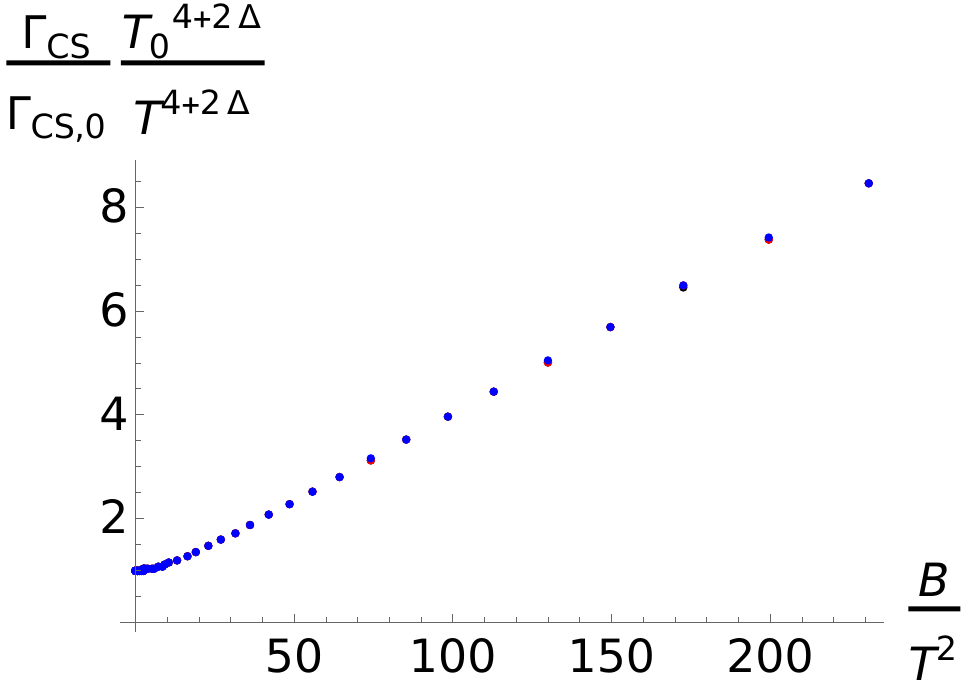} \includegraphics[width=0.49\linewidth]{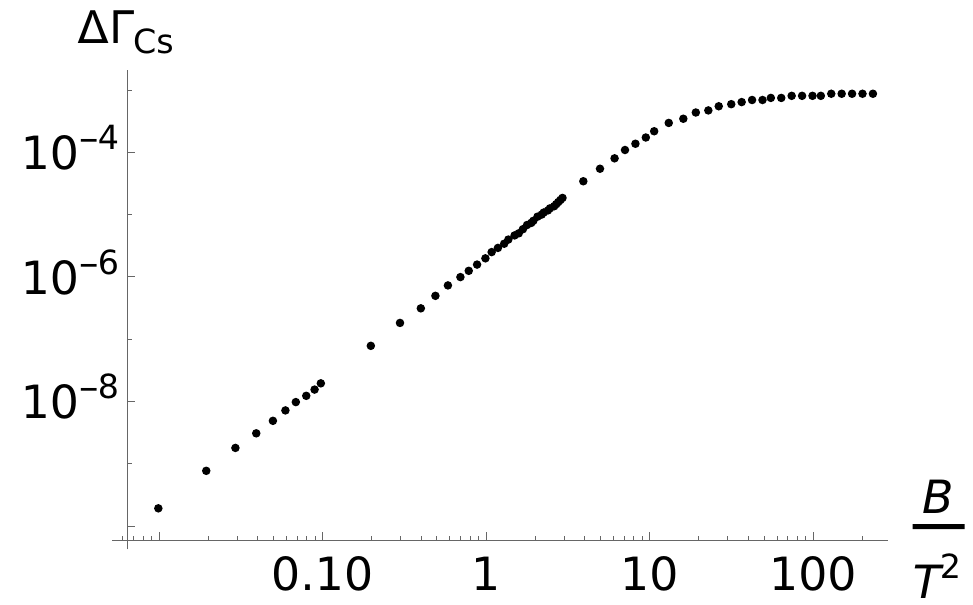}
    \caption{Left: Dimensionless Chern-Simons diffusion rate for $m_sL=0.04$ normalized to the dimensionless Chern-Simons rate at $B=0$ for $\alpha=2$ (blue), $\alpha=6/19$ (red) and $\alpha=0$ (black). Right: Effect of the Abelian anomaly by considering the quantity defined in \eqref{eq:appCSdiff} as function of the magnetic field.}\label{fig:CSapp2}
    \end{figure*}

In this section, we give some additional plots for the Chern-Simons diffusion rate discussed in section~\ref{sec:CSdiff}. 
The Chern-Simons diffusion rate for the decoupled axion in a strong magnetic field (neglecting anomalies and matter) was calculated in~\cite{Basar:2012gh}. Since we took a different approach to calculate the Chern-Simons diffusion rate, we connect our results to the calculation in~\cite{Basar:2012gh}. In the right panel of figure \ref{fig:CSapp2}, we show the enhancement of the Chern-Simons diffusion rate in a magnetic field compared to the zero $B$ case as a function of the magnetic field (for three different choices of $\alpha$). 
The scaling at small and large $B$ is given by
\begin{align}
      & \!\frac{\Gamma_\text{CS}}{\Gamma_\text{CS,B=0}}\frac{T_{B=0}^{4+2\Delta}}{T^{4+2\Delta}}\sim\begin{cases} 1+ 0.001 \left(\frac{B}{T^2}\right)^{2.0}\!\!,\text{ for } \frac{B}{T^2}\!\ll\! 1\\ 0.64 + 0.03 \left(\frac{B}{T^2}\right)^{1.0}\!\!,\text{ for } \frac{B}{T^2}\gg 1.\end{cases}
\end{align}
Both scalings are in agreement with~\cite{Basar:2012gh}.
Finally, to demonstrate the effect of the Abelian anomaly, we defined the quantity \begin{equation}
    \Delta \Gamma_\text{CS}=2\,\frac{\Gamma _{\text{CS},\alpha_1}-\Gamma_{\text{CS},\alpha_0}}{\Gamma _{\text{CS},\alpha_0}+\Gamma _{\text{CS},\alpha_1}}.\label{eq:appCSdiff}
\end{equation}
For the right panel in figure \ref{fig:CSapp2} we set $\alpha_1=6/19$ and $\alpha_0=0$.
We conclude that the Abelian anomaly enhances the Chern-Simons rate and the effect is becoming more pronounced for stronger magnetic fields. For weak magnetic fields the growth is exactly quadratic $\Delta\Gamma_\text{CS}\sim 2.01\cdot10^{-6} \,(B/T^2)^{2.00}$.

\bibliography{main}
\end{document}